\documentclass[aps,prl,superscriptaddress,twocolumn,showpacs,preprintnumbers,amsmath,amssymb]{revtex4}

\usepackage[dvips]{graphicx,color}
\usepackage{dcolumn}
\usepackage{bm}
\usepackage{ulem}
\usepackage{soul}
\begin{document}

\title{Inverse spin Hall effect in a closed loop circuit}

\author{Y.~Omori}
\affiliation{Institute for Solid State Physics, University of Tokyo, 5-1-5 Kashiwa-no-ha, Kashiwa, Chiba 277-8581, Japan}
\author{F.~Auvray}
\affiliation{Institute for Solid State Physics, University of Tokyo, 5-1-5 Kashiwa-no-ha, Kashiwa, Chiba 277-8581, Japan}
\author{T.~Wakamura}
\affiliation{Institute for Solid State Physics, University of Tokyo, 5-1-5 Kashiwa-no-ha, Kashiwa, Chiba 277-8581, Japan}
\author{Y.~Niimi}
\email{niimi@issp.u-tokyo.ac.jp}
\affiliation{Institute for Solid State Physics, University of Tokyo, 5-1-5 Kashiwa-no-ha, Kashiwa, Chiba 277-8581, Japan}
\author{A. Fert}
\affiliation{Unit\'{e} Mixte de Physique CNRS/Thales, 91767 Palaiseau France associ\'{e}e \`{a} l'Universit\'{e} de Paris-Sud, 91405 Orsay, France}
\author{Y.~Otani}
\affiliation{Institute for Solid State Physics, University of Tokyo, 5-1-5 Kashiwa-no-ha, Kashiwa, Chiba 277-8581, Japan}
\affiliation{RIKEN-CEMS, 2-1 Hirosawa, Wako, Saitama 351-0198, Japan}

\date{19 June 2014}

\begin{abstract}
We present measurements of inverse spin Hall effects (ISHEs) 
in which the conversion of a spin current into a charge current via the ISHE 
is detected not as a voltage in a standard open circuit but directly 
as the charge current generated in a closed loop. 
The method is applied to the ISHEs of Bi-doped Cu and Pt. 
The derived expression of ISHE for the loop structure can 
relate the charge current flowing into the loop 
to the spin Hall angle of the SHE material and the resistance of the loop.
\end{abstract}


\maketitle

Spintronic devices rely on the generation, manipulation, and detection 
of spin currents, flows of spin angular momentum~\cite{maekawa_review_book}. 
The spin Hall effect (SHE), originally predicted by Dyakonov and 
Perel~\cite{d.pl.1971} in 1971 and revived 
by Hirsch~\cite{hirsch_prl_1999} about thirty years later, 
is one of the ways to 
convert an electric charge current into a spin current. 
Since the spin current does not accompany the flow of charge causing 
energy dissipation, effective ways to produce the spin current have been 
intensively studied in the field of spintronics~\cite{seki_nat_mater_2008,liu_science_2012,ralph_apl_2012,n.prl.2012}. 
However, the spin current is not a conservative quantity 
but a diffusive flow. 
Thus, it cannot be directly observed but is measured via spin accumulation, 
and the spatial variation of spin accumulation is the spin current. 

Using the inverse process of SHE, i.e., inverse SHE (ISHE), 
a spin current can be converted into a charge current. 
In a standard open circuit measurement, 
the charge current induced by the ISHE gives rise to 
charge accumulation generating an electric 
voltage~\cite{valenzuela_nature_2006,saitoh_apl_2006,hoffmann_prl_2010}. 
Here we present experiments showing clearly 
that the ISHE can also be detected by the current in a closed loop.
The current in the loop can be measured as the voltage 
between two voltage probes in the loop.
Such a voltage drop in the nanoscale device is a clear 
evidence of the conversion of spin current into charge current, 
which has not directly been observed in previous experiments~\cite{valenzuela_nature_2006,saitoh_apl_2006,hoffmann_prl_2010}.
From the detailed analyses, it turns out that the amount of 
the charge current is determined by the spin Hall (SH) angle of SHE material 
and the resistance of the loop.

\begin{figure*}
\begin{center}
\includegraphics[width=17cm]{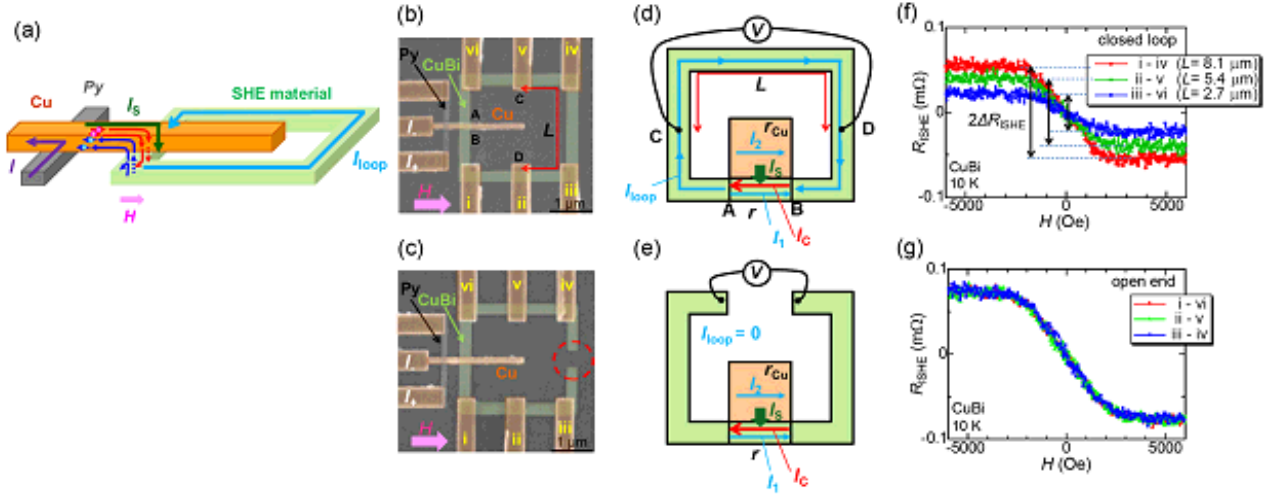}
\caption {(Color online) (a) Schematic of the measurement circuit. Spin-up and spin-down electrons $|e|$ ($|e|$ is the charge of the electron) denoted by spheres with arrows are absorbed into the SHE material and scattered to the same side by the ISHE. The induced electric current in the loop $I_{\rm loop}$ can be measured as a voltage drop. The magnetic field $H$ is applied along the hard direction of the Py wire. (b)-(e) Scanning electron micrographs and schematics of the 1D model for the closed loop circuit and open end circuit devices. The total length $L_{\rm M}$ of the loop [from A to B in (b) and (d)] is 11~$\mu$m. The distance $L$ between two voltage probes (for instance, between C and D) is defined as shown in (b) and (d). The broken circle is added in (c) to emphasize that the loop is cut. $r_{\rm Cu}$ and $r$ in (d) and (e) are the resistances of the Cu bridge and the SHE material only below the Cu bridge, respectively. $r_{\rm Cu}$ can be expressed as $xr/(1-x)$ using the shunting factor $x$~\cite{n.prl.2011}. $I_{\rm C}$ is the ISHE-induced charge current. $I_{1}$, $I_{2}$ and $I_{\rm loop}$ are ohmic currents accompanied with voltage drops. See the text for more details. (f), (g) The ISHE resistances $R_{\rm ISHE}$ of Cu$_{99.5}$Bi$_{0.5}$ with the closed loop and open end structures measured at 10~K at different voltage probe positions (i-vi). The amplitude of ISHE resistance, $\Delta R_{\rm ISHE}$, is defined as shown in (f).}
\label{sample}
\end{center}
\end{figure*}

Samples have been fabricated on a thermally oxidized silicon subtrate 
using electron beam lithography on polymethyl-methacrylate resist 
and subsequent lift-off process. 
We first patterned a 100~nm wide wire and deposited permalloy 
(Py; Ni$_{81}$Fe$_{19}$) by 30~nm. 
We also patterned a closed loop or an open end shape 
next to the Py wire and deposited 
two different SHE materials, i.e., Cu$_{99.5}$Bi$_{0.5}$ and Pt, by 20~nm. 
The length of the loop $L_{\rm M}$ is 6, 11, and 19~$\mu$m, and 
the distance ($d$) between the Py wire and the SHE wire is 500~nm.
The Py and SHE wires were bridged by a 100~nm wide and 
100~nm thick Cu wire transferring the spin current. 
We also deposited six Cu electrodes (i-vi in Fig.~\ref{sample}) 
to measure the voltage ($V_{\rm ISHE}$) induced by the ISHE. 
Before the deposition of the Cu bridge and electrodes, 
a careful Ar ion etching was 
carried out for 30 seconds to obtain transparent interfaces 
between the SHE material and Cu as well as between Py and Cu.
The detailed sample dimensions and other characteristics 
are listed in Table~\ref{table1}.

\begin{table}
\caption{Dimensions (width $w$ and thickness $t$) of Py, Cu, Cu$_{99.5}$Bi$_{0.5}$ and Pt wires constituting the SHE devices. We also show the resistivity $\rho$, the spin diffusion length $\lambda$ and the SH angle $\alpha_{\rm H}$ measured at $T=10$~K. The suffix X represents each material (N, F or M). The index (1D or 3D) indicates that the paramter is obtained with the one-dimensional (1D) or three-dimensional (3D) model. The values of $\lambda_{\rm X}^{\rm 1D/3D}$ are taken from our previous papers~\cite{n.prl.2012,n.prl.2013}.}
\label{table1}
\begin{ruledtabular}
\begin{tabular}{ccccc}
Parameter & Py (F) & Cu (N) & Cu$_{99.5}$Bi$_{0.5}$ (M) & Pt (M) \\
\hline
$w_{\rm X}$ (nm) & 100 & 100 & 250 & 200 \\
$t_{\rm X}$ (nm) & 30 & 100 & 20 & 20 \\
$\rho_{\rm X}$ ($\mu\Omega\cdot$cm) & 19 & 1.5 & 11 & 14 \\
$\lambda_{\rm X}^{\rm 1D/3D}$ (nm) & 5 & 1300 & $32/45$ & $11/10$ \\
$\alpha_{\rm H}^{\rm 1D}$ &  &  & $-0.11(\pm 0.01)$ & $0.014$ \\
$\alpha_{\rm H}^{\rm 3D}$ &  &  & $-0.22(\pm 0.02)$ & $0.021$ \\
\end{tabular}
\end{ruledtabular}
\end{table}

In Fig.~\ref{sample}(a), we show the principle of ISHE 
in an SHE ring using the spin absorption 
method~\cite{n.prl.2012, n.prl.2011, m.prb.2011, n.prb.2014, n.prl.2013}.
When the electric current $I$ flows from the Py wire to 
the left side of the Cu wire, 
the resulting spin accumulation induces a pure spin current ($I_{\rm S}$) 
on the right side of the Cu wire. 
The pure spin current is preferentially absorbed into 
the Cu$_{99.5}$Bi$_{0.5}$ or Pt ring. 
The opposite spin-up and spin-down electrons composing the absorbed pure 
spin current are deflected to the same direction by the ISHE. 
As we will detail later on, this conversion occurs only below the 
Cu bridge.

In the present work, we prepared two types of samples, 
i.e., closed loop circuit and open end one, as shown in 
scanning electron microscopy images in Figs.~\ref{sample}(b) 
and \ref{sample}(c) and in schematics 
in Figs.~\ref{sample}(d) and \ref{sample}(e).
Figures~\ref{sample}(f) and \ref{sample}(g) show ISHE resistances 
$R_{\rm ISHE}$ ($\equiv V_{\rm ISHE}/I$) of Cu$_{99.5}$Bi$_{0.5}$ 
measured with the closed loop and open end circuits, respectively.
$R_{\rm ISHE}$ increases with increasing the magnetic field $H$
and is saturated above 2000~Oe, 
which is the saturation field of magnetization of the Py wire, 
as already shown in our previous 
reports~\cite{m.prb.2011,n.prl.2011,n.prl.2012,n.prb.2014}. 
What is interesting to note is the amplitude of $R_{\rm ISHE}$, i.e., 
$\Delta R_{\rm ISHE}$. 
In the closed loop structure, $\Delta R_{\rm ISHE}$ depends on the 
distance $L$ between two voltage positions as shown in Fig.~\ref{sample}(f). 
For example, $\Delta R_{\rm ISHE}$ for $L=8.1$~$\mu$m is 
about three times larger than that for $L=2.7$~$\mu$m.
Such a voltage position dependence of $\Delta R_{\rm ISHE}$ has never 
been observed in the normal open end circuit [see Fig.~\ref{sample}(g)].

\begin{figure}
\includegraphics[width=8.5cm]{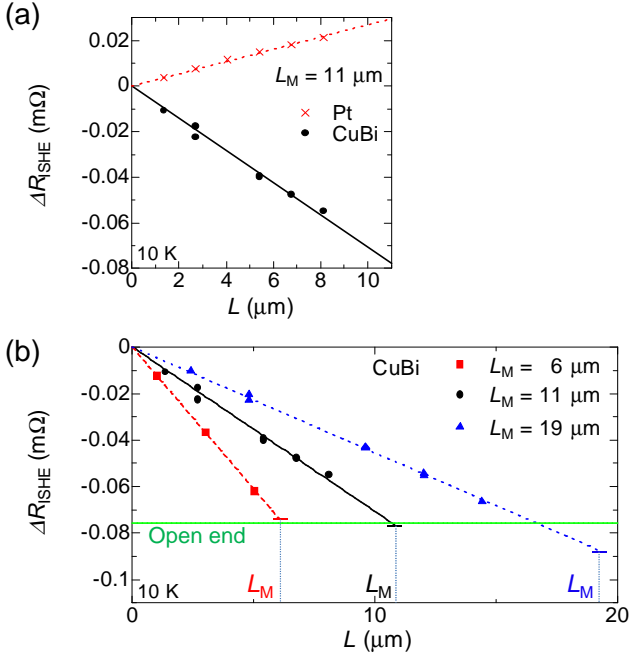}
\caption {(Color online) (a) $\Delta R_{\rm ISHE}$ of Cu$_{99.5}$Bi$_{0.5}$ ($\bullet$) and Pt ($\times$) rings with $L_{\rm M}=11$~$\mu$m as a function of $L$. The solid and broken lines are the linear fits. (b) The $L$ dependence of $\Delta R_{\rm ISHE}$ of Cu$_{99.5}$Bi$_{0.5}$ for different $L_{\rm M}$; $6~\mu$m (square), $11~\mu$m (circle), and $19~\mu$m (triangle). The broken, solid, and dotted lines are the linear fits. The endpoint of each line is coincident with $\Delta R_{\rm ISHE}$ measured with the open end circuit device.}
\label{result}
\end{figure}

To see clearly the relation between $\Delta R_{\rm ISHE}$ and $L$, 
we plot $\Delta R_{\rm ISHE}$ of the Cu$_{99.5}$Bi$_{0.5}$ and Pt rings
as a function of $L$ in Fig.~\ref{result}(a). 
Both of them show a linear dependence on $L$, but 
$\Delta R_{\rm ISHE}$ of Cu$_{99.5}$Bi$_{0.5}$ has a negative slope 
while that of Pt has a positive one. 
This is consistent with our previous works; 
the SH angle $\alpha_{\rm H}$ of Cu$_{99.5}$Bi$_{0.5}$ is 
negative~\cite{n.prl.2012} while 
that of Pt is positive~\cite{m.prb.2011}.
From the linear dependence of $\Delta R_{\rm ISHE}$ on $L$, 
we can conclude that an electric current ($I_{\rm loop}$) flows 
in the SHE ring and the direction of the current relies on 
the SH angle of the SHE material. 
From the slope of $\Delta R_{\rm ISHE}$ vs $L$ curve, 
we can estimate $I_{\rm loop}$ in the ring; 
$I_{\rm loop}=-0.18$~nA for Cu$_{99.5}$Bi$_{0.5}$ and 
$I_{\rm loop} =0.04~{\rm nA}$ for Pt 
when $I=0.58$~mA is applied. 
However, the obtained $I_{\rm loop}$ is too small when 
we simply consider the conversion of spin current into charge current 
via the ISHE.

We now formulate $I_{\rm loop}$ 
within the standard one-dimensional (1D)
spin diffusion model~\cite{t.prb.2003}.
As shown in Figs.~\ref{sample}(d) and \ref{sample}(e), 
the generated $I_{\rm S}$ is absorbed into the SHE ring because of 
its strong spin-orbit interaction. 
In this device structure, the absorbed pure spin current 
can be expressed as follows:
\begin{eqnarray}
\frac{\overline{I_{\rm S}}}{I}
=\frac{\lambda_{\rm M}}{t_{\rm M}}\frac{(1-e^{-t_{\rm M}/\lambda_{\rm M}})^{2}}{1-e^{-2t_{\rm M}/\lambda_{\rm M}}}\frac{2p_{\rm F}Q_{\rm F}e^{d/\lambda_{\rm N}}}{(2Q_{\rm M}+1)(2Q_{\rm F}+1)e^{2d/\lambda_{\rm N}}-1}
\label{Is}
\end{eqnarray}
where $\overline{I_{\rm S}}$ is defined as an average of pure spin current 
flowing vertically into the Cu$_{99.5}$Bi$_{0.5}$ or Pt ring. 
$R_{\rm N}, R_{\rm F}$ and $R_{\rm M}$ are respectively the spin resistances 
of Cu, Py and Cu$_{99.5}$Bi$_{0.5}$ or Pt, and 
$Q_{\rm F} = R_{\rm F}/R_{\rm N}$ and $Q_{\rm M} = R_{\rm M}/R_{\rm N}$. 
The spin resistance $R_{\rm X}$ is defined as 
$\rho_{\rm X} \lambda_{\rm X}/ (1-p^{2}_{\rm X})A_{\rm X}$, 
where $\rho_{\rm X}, \lambda_{\rm X}, p_{\rm X}$ and $A_{\rm X}$ are 
respectively the electrical resistivity, the spin diffusion length, 
the spin polarization, the effective cross sectional area involved 
in the equations of the 1D spin diffusion model and 
the suffix X represents each material (N, F or M). 
These values obtained at $T=10$~K are shown in Table~\ref{table1}.

The SH angle $\alpha_{\rm H}$ is the conversion rate 
between the pure spin current density 
$j_{\rm S} = \frac{\overline{I_{\rm S}}}{w_{\rm N}w_{\rm M}}$ 
and the charge current density 
$j_{\rm C} = \frac{I_{\rm C}}{t_{\rm M}w_{\rm M}}$. 
Thus, we obtain 
$I_{\rm C} = \frac{t_{\rm M}}{w_{\rm N}} \alpha_{\rm H} \overline{I_{\rm S}}$ 
[see the red arrow in Figs.~\ref{sample}(d) and \ref{sample}(e)]. 
For the open end case, no charge current 
flows in the circuit (i.e., $I_{\rm loop} = 0$). 
This means that the converted charge current $I_{\rm C}$
is compensated by $I_{1}$ and $I_{2}$ as depicted in 
Fig.~\ref{sample}(e): $I_{\rm C} = I_{1} + I_{2}$. 
Here $I_{1}$, $I_{2}$ and $I_{\rm loop}$ are 
currents flowing in the SHE material only below the Cu bridge, 
in the Cu bridge, and in the SHE material outside the Cu bridge, respectively. 
As can be seen in Fig.~\ref{sample}(g), 
$\Delta R_{\rm ISHE}$ has no voltage position dependence. 
Thus, the induced ISHE voltage $\Delta V_{\rm ISHE}$ can be written as follows:
\begin{eqnarray}
\Delta V_{\rm ISHE} = rI_{1} = \frac{xr}{1-x}I_{2} = \frac{1}{w_{\rm M}}\alpha_{\rm H}^{\rm 1D} x \overline{I_{\rm S}} \rho_{\rm M}
\label{1D_model}
\end{eqnarray}
where $r$ is the resistance of the SHE material below the Cu bridge 
[see the caption of Fig.~\ref{sample}], 
$\rho_{\rm M}$ is the longitudinal resistivity of the SHE material, and 
$x \simeq 0.36$ is the shunting factor, originally introduced 
in Ref.~\cite{n.prl.2011}. 
This expression is indeed the same as Eq.~(\ref{1D_model}) 
in Ref.~\cite{n.prl.2011}.

In the loop structure, $I_{\rm loop}$ is not zero anymore 
as shown in Fig.~\ref{sample}(d) and can be expressed as: 
\begin{eqnarray}
I_{\rm loop} \approx \frac{t_{\rm M}}{L_{\rm M}}\alpha_{\rm H}^{\rm 1D} x \overline{I_{\rm S}}. 
\label{1D_ring}
\end{eqnarray}
To obtain Eq.~(\ref{1D_ring}), we assume $L_{\rm M} \gg w_{\rm N}$. 
From this equation, we can explain that $I_{\rm loop}$ depends 
not only on $\alpha_{\rm H}^{\rm 1D}$ but also on $L_{\rm M}$. 
In addition, $\Delta V_{\rm ISHE}$ 
at the endpoint of the loop ($L = L_{\rm M}$) 
becomes Eq.~(\ref{1D_model}) since it is a product of 
$I_{\rm loop}$ and the resistance of the loop. 
As $I_{\rm loop}$ is already obtained from Fig.~\ref{result}, 
we can estimate $\alpha_{\rm H}^{\rm 1D}$ with Eq.~(\ref{1D_ring}). 
We note that unlike the case of the lateral spin valve 
device~\cite{n.prl.2012,n.prl.2011,m.prb.2011,n.prb.2014}, 
$\lambda_{\rm M}^{\rm 1D}$, which is needed to obtain $\overline{I_{\rm S}}$, 
cannot be directly determined on the same device. 
Therefore, as shown in Table~\ref{table1},
we have referred to $\lambda_{\rm M}^{\rm 1D}$ reported in our 
previous works~\cite{n.prl.2012,n.prl.2013} 
to obtain $\alpha_{\rm H}^{\rm 1D}$; 
$\alpha_{\rm H}^{\rm 1D}=-0.11$ for Cu$_{99.5}$Bi$_{0.5}$ 
and $\alpha_{\rm H}^{\rm 1D}=0.014$ for Pt. 
These results are consistent with 
our previous works~\cite{n.prl.2012,n.prl.2011,m.prb.2011,n.prb.2014}.

\begin{figure}
\includegraphics[width=8.5cm]{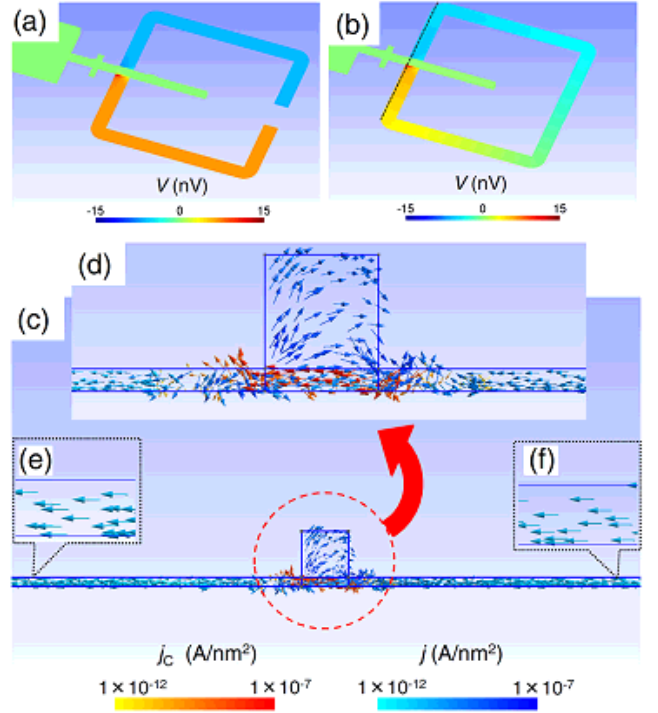}
\caption {(Color online) (a), (b) $V_{\rm ISHE}$ of Pt for (a) the open end circuit and (b) the closed loop circuit calculated with SpinFlow~3D. The vertical line in (b) corresponds to the cross sectional image shown in (c). (c) Calculated current densities generated by the ISHE $j_{\rm C}$ (red arrows) and induced by the electric field $j$ (blue arrows) for the Pt ring device. $j_{\rm C}$ and $j$ correspond to $I_{\rm C}$ and $I_{1}+I_{2}+I_{\rm loop}$ in the 1D model, respectively. The amplitude of the current density is expressed on color scale. (d)-(f) Closeup views of (c) in the vicinity of the Cu/Pt interface (d), at the left side of the ring (e), and at the right side of the ring (f).}
\label{discussion}
\end{figure}

So far, we have fixed $L_{\rm M}(=11~\mu{\rm m})$. 
To check Eq.~(\ref{1D_ring}), 
we prepared three different $L_{\rm M}$ rings [see Fig.~\ref{result}(b)]. 
As expected, $I_{\rm loop}$ depends on $L_{\rm M}$:
for instance, $I_{\rm loop}$ becomes $-0.31$ nA, 
for the Cu$_{99.5}$Bi$_{0.5}$ ring with $L_{\rm M}=6$~$\mu$m. 
Moreover, $\Delta R_{\rm ISHE}$ at the endpoint of the loop 
($L = L_{\rm M}$) coincides with 
$\Delta R_{\rm ISHE}$ of the open end circuit, 
as can be seen in Fig.~\ref{result}(b). 
This result also supports the above consideration.

To confirm our 1D analysis, we have also performed 
three-dimensional (3D) analyses 
for $V_{\rm ISHE}$ and $I_{\rm loop}$ 
using SpinFlow~3D based on the Valet-Fert 
formalism~\cite{n.prl.2012, v.prb.1993}. 
In the case of the open circuit device, 
the distribution of $V_{\rm ISHE}$ 
is homogeneous both for the left and right sides, and 
the difference between the two sides corresponds to $\Delta V_{\rm ISHE}$, 
as shown in Fig.~\ref{discussion}(a). 
This result also indicates that $I_{\rm loop}=0$.
When the SHE material has a closed loop structure, on the other hand, 
$V_{\rm ISHE}$ gradually changes from the positive value (on the left side) 
to the negative one (on the right side), but $\Delta V_{\rm ISHE}$ 
obtained at the two edges next to the Cu bridge is the same as that 
for the open end circuit [see Fig.~\ref{discussion}(b)]. 
This is consistent with the experimental result 
shown in Fig.~\ref{result}(b). 
To prove that the charge current flows in the ring, 
we show the charge current density distribution calculated with SpinFlow~3D 
in Figs.~\ref{discussion}(c)-\ref{discussion}(f). 
As in the case of the 1D model, 
$j_{\rm C}$ is compensated by $j$ below the Cu bridge, but 
since the SHE material has a ring shape, 
the small leakage current flows in the ring.
From the leakage current, 
we can also estimate $\alpha_{\rm H}^{\rm 3D}$ by 
using $\lambda_{\rm M}^{\rm 3D}$ shown in Table~\ref{table1}; 
$\alpha_{\rm H}^{\rm 3D}=-0.22$ for Cu$_{99.5}$Bi$_{0.5}$ 
and $\alpha_{\rm H}^{\rm 1D}=0.021$ for Pt. 
These are again consistent with 
our previous works~\cite{n.prl.2012,n.prb.2014}.

In Fig.~\ref{discussion}, 
we selected Pt as an SHE material 
to show the current distribution in the ring.
We obtain the similar current distribution for CuBi 
but with the opposite direction compared to Pt. 
However, the current distribution below the Cu bridge 
is much more complicated to see than that for Pt. 
This originates from the spreading of the spin accumulation 
at the side edges of the CuBi ring 
since $\lambda_{\rm CuBi}$ is larger than 
$t_{\rm CuBi}$~\cite{n.prl.2012}.  

Finally, we discuss how $I_{1}$, $I_{2}$ and $I_{\rm loop}$ are generated 
in the closed loop circuit and how $I_{\rm loop}$ can be utilized. 
As shown in Fig.~\ref{sample}(d), 
the induced $I_{\rm S}$ is injected into the SHE material from the Cu bridge. 
We note here that 
the conversion of $I_{\rm S}$ into $I_{\rm C}$ occurs only 
at the Cu/SHE-material junction. 
Thus, electrons converted from $I_{\rm S}$ lose the driving force once 
they go out from the junction, and are accumulated to one side. 
This induces the electric field in the SHE wire. 
For the open end circuit, the accumulated electrons are balanced 
with the electric field, as in the case of the normal Hall effect. 
As a result, $I_{\rm C}$ is cancelled out by $I_{1}$ and $I_{2}$, and 
the electric voltage $\Delta V_{\rm ISHE}$ can be measured. 
When the circuit is closed, 
the electric field also induces $I_{\rm loop}$ in the loop. 
This $I_{\rm loop}$ is essentially different from a current due to 
electromotive force induced by an alternating magnetic field through a ring. 
Although we have used the ac lock-in technique to obtain $I_{\rm loop}$, 
it is in principle a dc current. 
The present result clearly shows that 
by flowing $I$ from Py to Cu non-locally, 
another steady current can be induced in a mesoscopic ring 
with a large $\alpha_{\rm H}$ via the ISHE. 

As discussed above, since $I_{\rm loop}$ is proportional to 
$t_{\rm M}/L_{\rm M}$, it is of the order of 1~nA at the moment. 
By optimizing the device structure, the value can be enhanced 
by a factor of ten. 
Furthermore, by replacing 
a part of SHE ring with a superconductor, 
we could obtain a smaller resistance of the loop 
compared to the resistance of the shunt (Cu bridge). 
In such a case, $I_{\rm loop}$ should be enhanced closer to $I_{\rm C}$.
It is known that when there is a magnetic flux $\phi$ 
threading such a mesoscopic ring, 
a small persistent current (of the order of 1~nA) 
is induced in the ring and shows an oscillation 
with a period of $\phi_{0} =h/e$ where $h$ is 
the Plank constant~\cite{levy_prl_1990,webb_prl_1991,mailly_prl_1993}.
In the same way as a dc-SQUID magnetometer, 
by arranging the SHE ring on top of a sample and 
measuring a current in the ring precisely, one could observe a modulation of 
$I_{\rm loop}$ and thus extract the magnetization of the sample. 

In summary, we have measured the ISHEs of CuBi and Pt by means of 
the spin absorption technique in both open end and closed loop circuits, 
and found that the electric current can be obtained
only in the closed loop. 
The detected current depends on the spin Hall angle of the SHE material and 
also on the resistance of the loop.
It is commonly considered that the ISHE converts a spin current 
into a charge current, but 
this is the first observation of the converted charge current 
via the ISHE of CuBi or Pt.
The amount of the charge current can be quantitatively 
explained by our spin transport models. 

We acknowledge helpful discussions with T.~Kato. 
We would also like to thank Y.~Iye and S.~Katsumoto 
for the use of the lithography facilities. 
This work was supported by KAKENHI 
(Grant No. 24740217 and 23244071) 
and by Foundation of Advanced Technology Institute.

\end{document}